\documentclass[journal=langd5,manuscript=article]{achemso}

\usepackage[version=3]{mhchem} 
\usepackage{bm}
\usepackage{amsmath,amssymb}
\usepackage{txfonts}
\usepackage{type1cm}
\usepackage{subfigure} 
\usepackage{color}
\usepackage{ulem}

\author{Masao Iwamatsu}
\email{iwamatsu@ph.ns.tcu.ac.jp}
\phone{+81 (0)3 5705 0104 ext. 2382}
\fax{+81 (0)3 5707 2222}
\affiliation[Tokyo City University]
{Department of Physics, Faculty of Liberal Arts and Sciences, Tokyo City University, Setagaya-ku, Tokyo 158-8557, JAPAN}

\title[Free-energy barrier]{Free-energy barrier of filling a spherical cavity in the presence of line tension: Implication to the energy barrier between the Cassie and Wenzel state  on a superhydrophobic surface with spherical cavities}

\abbreviations{IR,NMR,UV}

\begin{document}
\begin{abstract}
The free-energy barrier of filling a spherical cavity having an inner wall of various wettabiities is studied.  The morphology and free energy of a lens-shaped droplet are determined from the minimum of the free energy.  The effect of line tension on the free energy is also studied.  Then, the equilibrium contact angle of the droplet is determined from the generalized Young's equation.  By increasing the droplet volume within the spherical cavity, the droplet morphology changes from spherical with an equilibrium contact angle of $180^{\circ}$ to a lens with a convex meniscus, where the morphological complete drying transition occurs.  By further increasing the droplet volume, the meniscus changes from convex to concave. Then, the lens-shaped droplet with concave meniscus spreads over the whole inner wall resulting in an equilibrium contact angle of $0^{\circ}$ to leave a spherical bubble, where the morphological complete wetting transition occurs.  Finally, the whole cavity is filled with liquid.  The free energy shows a barrier from complete drying to complete wetting as a function of droplet volume, which corresponds to the energy barrier between the Cassie and Wenzel state of the superhydrophobic surface with spherical cavities.  The free-energy maximum occurs when the meniscus of the droplet becomes flat and it is given by an analytic formula.  The effect of line tension is expressed by the scaled line tension, and this effect is largest at the free-energy maximum.  The positive line tension increases the free-energy maximum, which thus, increases the stability of the Cassie superhydrophobic state, whereas the negative line tension destabilizes the superhydrophobic state.     

\end{abstract}

\section{\label{sec:sce1}Introduction}

Liquid confinement within various micro- and nano-cavities is a ubiquitous phenomenon, that occurs in material processing such as heterogeneous liquid or bubble nucleation (condensation)~\cite{Kelton2010,Yarom2015,Cooper2008} or surface sciences such as the wetting and spreading of liquids on structured surfaces~\cite{Bormashenko2013a,Tuteja2007} or on skin tissue or various types of membranes~\cite{Song2014,Li2008}.  In some of these problems, knowledge of the free energy and contact angle of a lens-shaped droplet placed on the bottom of a spherical cavity is crucial for understanding the physics behind these phenomena. 

When a droplet wets a substrate, however, the line tension~\cite{Bormashenko2013a,Gibbs1906,deGennes1985,Bonn2009,Weijs2011} at the three-phase contact line should play some role in determining the morphology and the free energy of the droplet.  The line tension is particularly important for nanoscale droplets~\cite{Bonn2009,Weijs2011,Pompe2000,Wang2001,Checco2003,Schimmele2007,Maheshwari2016}.  In our previous papers~\cite{Iwamatsu2015a,Iwamatsu2016a}, we pointed out that the line tension plays a fundamental role in the stability of nonvolatile lens-shaped droplets placed on the bottom of a spherical cavity.

This problem is closely related to the stability of a super hydrophobic surface.  The free energy of the droplet inside the cavity will provide information on the energy barrier that is needed to induce the empty--filled transition called Cassie--Wenzel transition, which destroys the hydrophobicity~\cite{Marmur2008,Sheng2007,Whyman2011,Savoy2012,Giacomello2012b,Verho2012,Bormashenko2013b,Papadopoulos2013,Checco2014}  if the imbibition into all pores in parallel is a reasonable assumption. In particular, Abdelsalam et al.~\cite{Abdelsalam2005} have succeeded in fabricating hydrophobic substrates with spherical cavities~\cite{Abdelsalam2005} from hydrophilic gold surface, where the air-filled empty cavity is more stable than the filled cavity and the intrusion of a liquid into the cavity is prohibited.  Subsequently, various theoretical model calculations for surfaces with spherical cavities have conducted~\cite{Marmur2008,Whyman2011,Bormashenko2013b,Patankar1999}.  However, they did not take into account the effect of line tension except the work by Bormashenko and Whyman~\cite{Bormashenko2013b}, who showed that the line tension can be important to determine the energy barrier of Cassie-Wenzel transition.  However, they did not aware of possibility of wetting and drying transitions~\cite{Iwamatsu2016a} within the spherical cavity because they did not consider the  changing meniscus curvature with volume.

In this paper, we will consider the evolution of the free energy and morphology of a droplet placed on the bottom of a spherical cavity as a function of droplet volume.  We extend our previous work~\cite{Iwamatsu2016a} and consider the line-tension effects on the free energy and morphology of a droplet, by which we take into account the wetting and drying transitions as well as the changing meniscus.  We will use the terminology "hydrophobic" and "hydrophilic" in this paper, although the terms "hygrophobic" and "hygrophilic" would be more general~\cite{Marmur2008}.   

We find that the free-energy barrier depends strongly on the wettability of the substrate characterized by the intrinsic Young's contact angle.   Naturally, the free-energy barrier increases for more hydrophobic surfaces (larger Young's contact angle).  The maximum of the free-energy barrier is given by an analytic formula.  The effect of line tension is greatest at the free-energy barrier maximum, and its effect is characterized solely by the scaled line tension.  This result supports the conclusion reached by Bormashenko and Whyman~\cite{Bormashenko2013b} using a more  simplistic model calculation.  In addition, we find a morphological complete drying transition from a spherical droplet to a lens-shaped droplet when the droplet volume is small and a complete wetting transition when the droplet volume is large in accordance with the prediction of our previous paper~\cite{Iwamatsu2016a}.  More specifically, the droplet becomes spherical in the complete drying transition, and it spreads over the whole wall of the inner substrate of the cavity to leave a spherical bubble in the complete wetting transition.   In this paper, we will use the term "wetting/drying transitions," although they are not thermodynamic phase transitions because the size of a liquid droplet is finite.  In section 2, we will discuss the morphology of a droplet within a cavity in the presence of line tension using the formula for the free energy from previous paper~\cite{Iwamatsu2016a}.  In section 3, we will discuss the free-energy barrier of the Cassie-Wenzel transition when the droplet volume is altered, which includes a discussion of the stability of the superhydrophobic Cassie state.  Our conclusions are given in section 4.

\section{\label{sec:sec2}Morphology of a droplet within a cavity}

\begin{figure}[htbp]
\begin{center}
\includegraphics[width=0.70\linewidth]{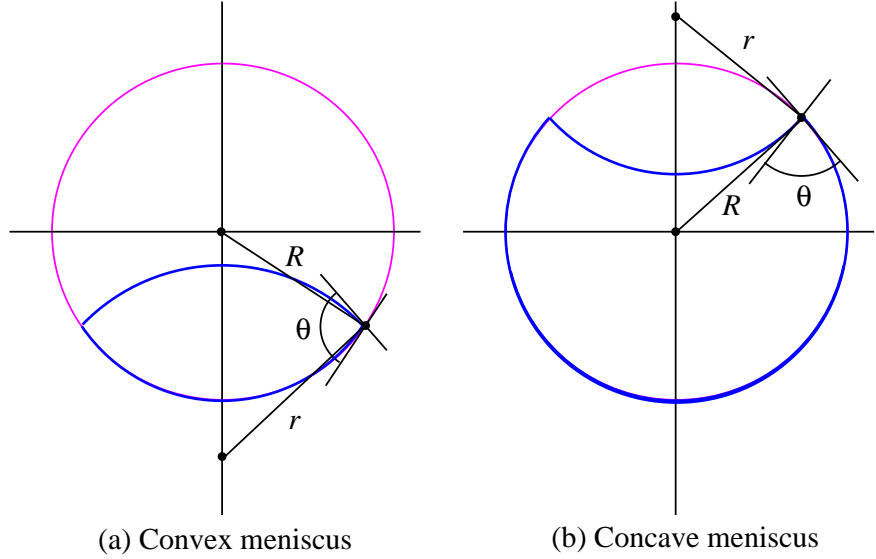}
\caption{
(a) A lens-shaped droplet of contact angle $\theta$ with a convex meniscus. Note that this droplet is equivalent to the lens-shaped bubble with a convex meniscus in (b). (b) A droplet with a concave meniscus, which is regarded as a bubble. As the droplet volume increases, the droplet morphology changes from convex to concave for the fixed contact angle $\theta$.  
   }
\label{fig:B1}
\end{center}
\end{figure}

In our previous work~\cite{Iwamatsu2015a,Iwamatsu2016a}, we studied the morphology of a lens-shaped droplet  of the non-volatile liquid placed on an inner bottom of a spherical cavity induced by the presence of line tension, as shown in Fig.~\ref{fig:B1}.  Here, we extend our previous study and focus on the evolution of droplet morphology and its free energy with respect to the liquid volume.  In this section, we will present a minimum formula and provide the background for our discussion of the volume-dependent free energy.  The detailed derivation and explanation have already been given in our previous publication~\cite{Iwamatsu2016a}.

We consider a droplet with a spherical surface of radius $r$ and contact angle $\theta$ placed on the bottom of a spherical cavity of radius $R$.  For each fixed droplet volume $V$, we consider the most stable structure of the droplet. We can then study the evolution of the droplet morphology and its free energy when the liquid volume is altered by the injection or extraction of a liquid.  We use the so-called {\it capillary model}, where the structure and width of the interfaces are neglected and the liquid-vapor, liquid-solid, and solid-vapor interactions are accounted for by the curvature-independent surface tensions. 

Although we consider the free-energy barrier of filling a single spherical cavity by a liquid, the results can be used to study the stability of the Cassie superhydrophobic state and the Cassie--Wenzel transition because a superhydrophobic substrate with a re-entrant structure can be modeled by a substrate having spherical pores~\cite{Bormashenko2013b,Marmur2008}.  In order to study the intrusion of a liquid onto the re-entrant substrate, we have to consider the situation where the liquid is injected from the top of the pore instead of the bottom of the pore, and so we have to turn Fig.~\ref{fig:B1} upside down.

The meniscus of a droplet can be convex, concave, or flat depending on the magnitude of the contact angle $\theta$.  The contact angle $\theta_{\infty}$ for a flat substrate is determined from the implicit equation
\begin{equation}
V=\frac{\pi}{3}\left(2-3\cos\theta_{\infty}+\cos^{3}\theta_{\infty} \right)R^{3}
\label{eq:B1}
\end{equation}
for the fixed droplet volume $V$.  

\begin{figure}[htbp]
\begin{center}
\includegraphics[width=0.70\linewidth]{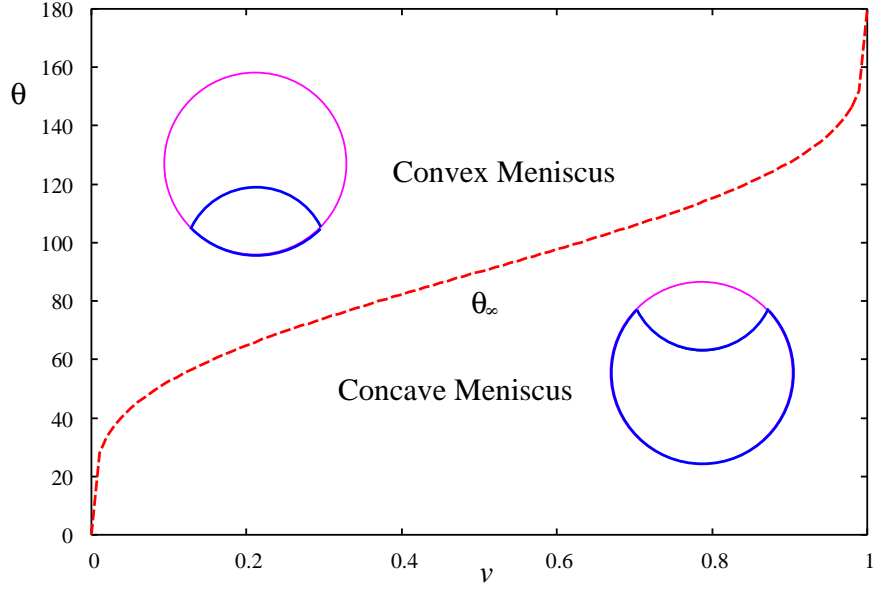}
\caption{
The contact angle $\theta_{\infty}$ for a flat meniscus as a function of the droplet's dimensionless volume $v$.  The  left-hand side of the curve corresponds to the convex meniscus and the right-hand side to the concave meniscus.  The meniscus changes from convex to concave at the volume where $\theta=\theta_{\infty}$ is satisfied as the droplet size is increased when the contact angle $\theta$ is fixed.  
 }
\label{fig:B2}
\end{center}
\end{figure}

In Fig.~\ref{fig:B2}, we show the contact angle $\theta_{\infty}$ as a function of the droplet's dimensionless volume $v$ defined by
\begin{equation}
V=\frac{4\pi}{3}R^{3}v.
\label{eq:B2}
\end{equation}
For a fixed contact angle $\theta$, the meniscus changes from convex to concave as the dimensionless volume $v$ is increased.  Furthermore, the meniscus is mostly convex (droplet) when the substrate is hydrophobic ($\theta>90^{\circ}$) and it is mostly concave (bubble) when the substrate is hydrophilic ($\theta<90^{\circ}$).  The purpose of this paper is to study the transformation of the droplet (bubble) morphology and free energy induced by the action of line tension  when the droplet volume is altered.  

In order to determine the most stable droplet shape and its free energy, we have to find the morphology that minimizes the free energy of a droplet in the capillary model given by
\begin{equation}
F=\sigma_{\rm lv}A_{\rm lv}+\Delta\sigma A_{\rm sl}+\tau L
\label{eq:B3}
\end{equation}
and
\begin{equation}
\Delta\sigma = \sigma_{\rm sl}-\sigma_{\rm sv}=-\sigma_{\rm lv}\cos\theta_{\rm Y},
\label{eq:B4}
\end{equation}
where $A_{\rm lv}$ and $A_{\rm sl}$ are the surface areas of the liquid-vapor and liquid-solid (substrate) interfaces, respectively, and $\sigma_{\rm lv}$ and $\sigma_{\rm sl}$ are their respective surface tensions.  The surface tension $\sigma_{\rm sv}$ is the solid-vapor surface tension when the substrate is not covered by the wetting layer of the liquid.  The wettability (hydrophilicity and hydrophobicity) of the substrate of the cavity is characterized by the intrinsic Young's contact angle $\theta_{\rm Y}$ in Eq.~(\ref{eq:B4}), which is known as the classical Young's equation~\cite{Young1805}.  The effect of the line tension $\tau$ is given by the last term in Eq.~(\ref{eq:B3}), where $L$ denotes the length of the three-phase contact line.  When the line tension is positive ($\tau>0$), the droplet tends to minimize or even eliminate the line length  $L$ to achieve a lower free energy $F$.  When the line tension is negative ($\tau<0$), the droplet tends to maximize the line length  $L$.

Within the capillary approximation, the free energy (Eq.~(\ref{eq:B3})) of a lens-shaped droplet with a convex meniscus is given by~\cite{Iwamatsu2016a}
\begin{equation}
F=4\pi R^{2}\sigma_{\rm lv}f\left(\rho,\theta\right),
\label{eq:B5}
\end{equation}
with
\begin{eqnarray}
f\left(\rho,\theta\right)
&=& \rho\frac{1-\left(\xi-\rho\right)^{2}}{4\xi}-\cos\theta_{\rm Y}\frac{\rho^{2}-\left(\xi-1\right)^{2}}{4\xi}+\frac{\tilde{\tau}\rho}{2\xi}\sin\theta, \nonumber \\
&&\;\;\;\;\;\;\;\;(\theta>\theta_{\infty}, \mbox{Convex}),
\label{eq:B6}
\end{eqnarray}
where
\begin{equation}
\tilde{\tau}=\frac{\tau}{\sigma_{\rm lv}R}
\label{eq:B7}
\end{equation}
is the scaled line tension, 
\begin{equation}
\xi=\sqrt{1+\rho^{2}+2\rho\cos\theta},
\label{eq:B8}
\end{equation}
and
\begin{equation}
\rho=\frac{r}{R}
\label{eq:B9}
\end{equation}
is the size parameter of the droplet.  Similarly, the droplet volume is given by~\cite{Iwamatsu2016a}
\begin{equation}
V=\frac{4\pi}{3}R^{3}\omega\left(\rho,\theta\right),
\label{eq:B10}
\end{equation}
with
\begin{eqnarray}
\omega\left(\rho,\theta\right)&=&\frac{1}{16\xi}\left(\xi-1-\rho\right)^{2} \nonumber \\
&&\times\left[3\left(1-\rho\right)^{2}-2\xi\left(1+\rho\right)-\xi^{2}\right],\;\;\;  \nonumber \\
&&(\theta>\theta_{\infty}, \mbox{Convex}).
\label{eq:B11}
\end{eqnarray}
Then, the radius $r$ or the size parameter $\rho$ of the droplet will be a function of the contact angle $\theta$ for a given dimensionless volume $v$ such that
\begin{equation}
\omega\left(\rho,\theta\right)=v\;\;\;\;\Rightarrow\;\;\;\;\rho=\rho\left(\theta\right),
\label{eq:B12}
\end{equation}
and the free energy in Eq.~(\ref{eq:B6}) becomes a function of the contact angle $f=f\left(\theta\right)$ for a fixed dimensionless volume $v$.  

Although we are discussing the free-energy barrier, we consider the Helmholtz free energy of a fixed volume instead of the Gibbs free energy~\cite{Iwamatsu2015a} or grand potential~\cite{Lefevre2004} of the variable volume, which are used to study nucleation problems.  This is because we consider a nonvolatile liquid so that the volume is fixed at the equilibrium.  The droplet volume is controlled not by the vapor pressure but by the forced injection or extraction of the liquid from outside the cavity.  The free-energy barrier should be overcome by an external force instead of by thermal fluctuation.

The free energy and droplet volume for a concave meniscus are obtained simply by changing the sign of $\rho$ and $\xi$ in Eqs.~(\ref{eq:B6}) and (\ref{eq:B11}) as follows:
\begin{equation}
\rho \rightarrow -\rho,\;\;\;\; \xi \rightarrow -\zeta,
\label{eq:B13}
\end{equation}
with
\begin{equation}
\zeta=\sqrt{1+\rho^{2}-2\rho\cos\theta}.
\label{eq:B14}
\end{equation}
Therefore, we will only present the formulas for the convex meniscus from now on for brevity.  The formulas for the concave meniscus can be easily derived using the transformation in Eqs.~(\ref{eq:B13}) and (\ref{eq:B14}).

The equilibrium contact angle $\theta_{\rm e}$ is determined by minimizing the free energy $f=f\left(\theta\right)$  in Eq.~(\ref{eq:B6}) with respect to the contact angle $\theta$ under the condition of a constant dimensionless volume $v$, which gives the generalized Young's equation~\cite{Iwamatsu2015a}
\begin{eqnarray}
&&
\sigma_{\rm sl}-\sigma_{\rm lv}=-\sigma_{\rm lv}\left(\cos\theta_{\rm e}+\tilde{\tau}\frac{1+\rho_{\rm e}\cos\theta_{\rm e}}{\rho_{\rm e}\sin\theta_{\rm e}}\right), \nonumber \\
&&\;\;\;\;\;\;\;\;\;(\theta_{\rm e}>\theta_{\infty}, \mbox{Convex}),
\label{eq:B15}
\end{eqnarray}
or
\begin{eqnarray}
&&
\cos\theta_{\rm Y}=\cos\theta_{\rm e}+\tilde{\tau}\frac{1+\rho_{\rm e}\cos\theta_{\rm e}}{\rho_{\rm e}\sin\theta_{\rm e}}, \nonumber \\
&&\;\;\;\;\;\;\;\;\;(\theta_{\rm e}>\theta_{\infty}, \mbox{Convex}),
\label{eq:B16}
\end{eqnarray}
from Eq.~(\ref{eq:B4}), where $\rho_{\rm e}=\rho\left(\theta_{\rm e}\right)$.  Equation (\ref{eq:B15}) reduces to the classical Young's equation in Eq.~(\ref{eq:B4}) and $\theta_{\rm e}=\theta_{\rm Y}$ when $\tilde{\tau}=0$.  The corresponding minimized (extreme) free energy of a lens-shaped droplet is given by~\cite{Iwamatsu2016a}
\begin{equation}
F_{\rm lens}=4\pi R^{2}\sigma_{\rm lv}f_{\rm lens},
\label{eq:B17}
\end{equation}
with
\begin{eqnarray}
f_{\rm lens}&=&\frac{\left(1+\rho_{\rm e}-\xi_{\rm e}\right)^{2}\left(\cos\theta_{\rm e}+1+\xi_{\rm e}\right)}{4\xi_{\rm e}}
\nonumber \\
&&-\tilde{\tau}\frac{\left(1+\rho_{\rm e}\cos\theta_{\rm e}-\xi_{\rm e}\right)}{2\rho_{\rm e}\sin\theta_{\rm e}},\nonumber \\
&&(\theta_{\rm e}>\theta_{\infty}, \mbox{Convex}),
\label{eq:B18}
\end{eqnarray}
where $\xi_{\rm e}$ is given by Eq.~(\ref{eq:B8}) with $\rho$ and $\theta$ replaced by $\rho_{\rm e}=\rho\left(\theta_{\rm e}\right)$ and $\theta_{\rm e}$ is determined from Eq.~(\ref{eq:B15}).

\begin{figure}[htbp]
\begin{center}
\includegraphics[width=0.70\linewidth]{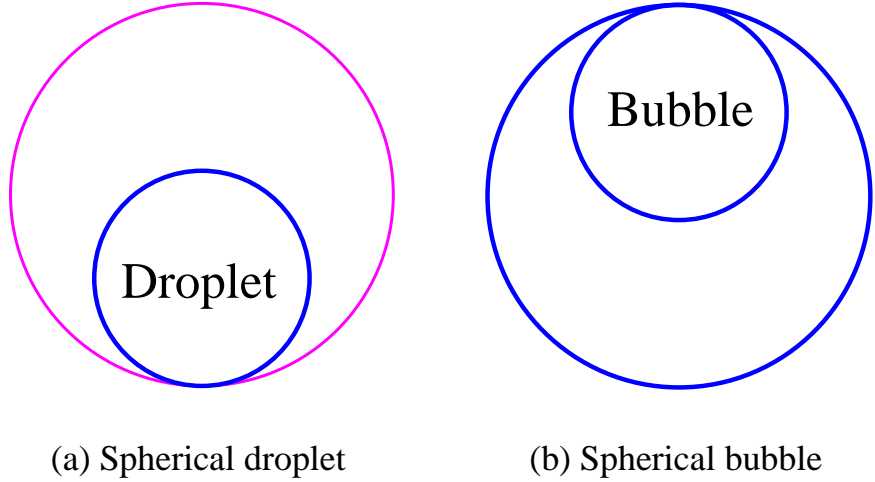}
\caption{
(a) A spherical droplet of contact angle $\theta_{\rm e}=180^{\circ}$ with convex meniscus (complete drying state).  (b) A spherical droplet of contact angle $\theta_{\rm e}=0^{\circ}$ with concave meniscus, which is regarded as a spherical bubble (complete wetting state). 
   }
\label{fig:B3}
\end{center}
\end{figure}

In our previous paper~\cite{Iwamatsu2016a}, we showed that the most stable structure is not only a lens-shaped droplet but also a spherical droplet with contact angle $\theta=180^{\circ}$ sitting on the bottom of a spherical cavity  [Fig.~\ref{fig:B3}(a)] and a droplet that is completely spread over the whole surface of a spherical cavity [Fig.~\ref{fig:B3}(b)] to form a spherical bubble with contact angle $\theta=0^{\circ}$ attached to the top wall of the cavity, as shown in Fig.~\ref{fig:B3}.   The free energy $F_{\rm drop}$ of a spherical droplet [Fig.~\ref{fig:B3}(a)] is given by the limit $\theta\rightarrow 180^{\circ}$ of Eq.~(\ref{eq:B6}), and is written as
\begin{equation}
F_{\rm drop}=4\pi R^{2}\sigma_{\rm lv}f_{\rm drop},
\label{eq:B19}
\end{equation}
where
\begin{equation}
f_{\rm drop}=\left(\rho_{\pi}\right)^{2}=\left(v\right)^{2/3},
\label{eq:B20}
\end{equation}
where $\rho_{\pi}=\rho\left(\theta=180^{\circ}\right)$.  By comparing the free energy $f_{\rm lens}$ of the lens-shaped droplet with $f_{\rm drop}$ of a spherical droplet, we can study the morphological transition between a lens-shaped droplet and a spherical droplet, which is the {\it drying transition} predicted on a flat substrate~\cite{Widom1995}.  Similarly, the free energy $F_{\rm bubble}$ of bubble is given by the $\theta\rightarrow 0^{\circ}$ limit of the free energy in Eq.~(\ref{eq:B6}) for the concave meniscus, which is given by
\begin{equation}
F_{\rm bubble}=4\pi R^{2}\sigma_{\rm lv}f_{\rm bubble},
\label{eq:B21}
\end{equation}
where
\begin{equation}
f_{\rm bubble}=\rho_{0}^{2}-\cos\theta_{\rm Y}
\label{eq:B22}
\end{equation}
and $\rho_{0}=\rho\left(\theta=0^{\circ}\right)$ is the size parameter when the contact angle is $\theta=0^{\circ}$.  By comparing the free energy $f_{\rm lens}$ of the lens-shaped droplet with $f_{\rm bubble}$ of the spherical bubble, we can study the morphological transition between a lens-shaped droplet and a spherical bubble, which might be termed the {\it wetting transition}.  

The relative stability of a spherical droplet and a spherical bubble is determined from $f_{\rm drop}=f_{\rm bubble}$, which leads to the condition for the Young's contact angle
\begin{equation}
\theta_{\rm Y,w}=\cos^{-1}\left[\left(1-v\right)^{2/3}-\left(v\right)^{2/3}\right]
\label{eq:B23}
\end{equation}
for a given dimensionless volume $v$, where we used
\begin{equation}
v=\left(\rho_{\pi}\right)^{3}=1-\left(\rho_{0}\right)^{3}
\label{eq:B24}
\end{equation}
derived from the condition of the conservation of droplet volume.

\begin{figure}[htbp]
\begin{center}
\includegraphics[width=0.70\linewidth]{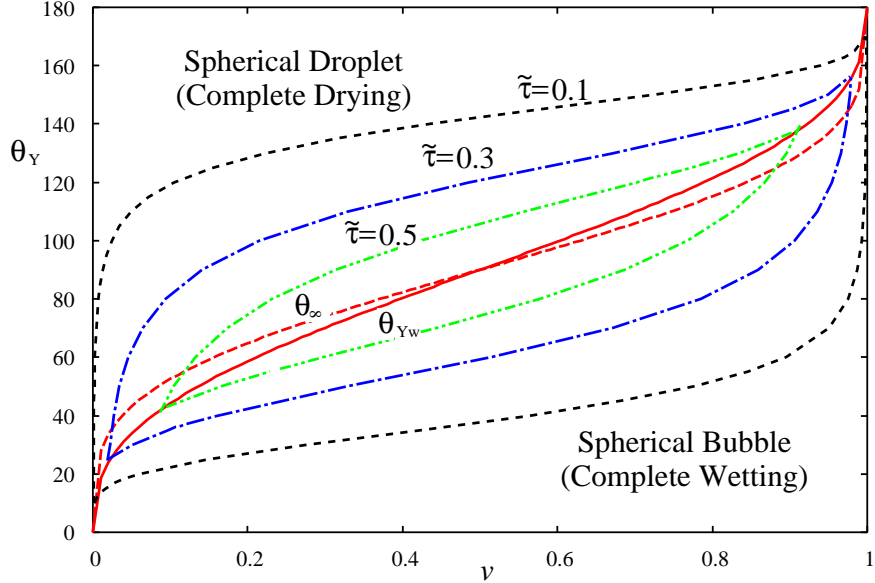}
\caption{
Morphological phase diagram of a droplet sitting on the bottom of a spherical cavity.  The morphological phase boundary between a lens-shaped droplet and a spherical droplet (drying transition, upper curves) and that between a lens-shaped droplet and a spherical bubble (wetting transition, lower curves) are shown.  In addition, the boundary $\theta_{\infty}$, where the meniscus becomes flat, and $\theta_{\rm Y,w}$, where the relative stability of a droplet and a bubble changes, are also shown. 
  }
\label{fig:B4}
\end{center}
\end{figure}

In Fig.~\ref{fig:B4}, we show the morphological phase boundaries derived from $f_{\rm drop}=f_{\rm lens}$ (drying transition) and  $f_{\rm bubble}=f_{\rm lens}$ (wetting transition).  First of all, the effect of line tension is represented by a non-dimensional scale line tension $\tilde{\tau}$ defined by Eq.~(\ref{eq:B7}).   A lens-shaped droplet is stable within a closed region sandwiched between an upper and lower curve, where a spherical droplet and a spherical bubble can only be metastable.  Outside of this closed region, either a spherical droplet (upper left region) or a spherical bubble (lower right region) is most stable.  The stability boundary $\theta_{\rm Y,w}$ between a spherical droplet and a spherical bubble, as well as the convex-concave boundary $\theta_{\infty}$, are also shown in Fig.~\ref{fig:B4}.   A spherical droplet (Fig.~\ref{fig:B3}(a)) is more stable than a spherical bubble (Fig.~\ref{fig:B3}(b)) if $\theta_{\rm Y}>\theta_{\rm Y,w}$ and vice versa.  Young's contact angle $\theta_{\rm Y,w}$ plays the role of a boundary between hydrophobicity and hydrophilicity of the concave spherical substrate. Note, however, that this boundary $\theta_{\rm Y,w}$ is determined solely by the droplet volume from Eq.~(\ref{eq:B23}).  In the next section, we will consider the evolution of the morphology and the free energy of a droplet placed on the bottom of a spherical cavity when the droplet volume is altered.

In order to observe this morphological transformation in Fig.~\ref{fig:B4}, the line tension must be as large as $\tilde{\tau}\sim 0.1$.  Suppose the liquid has a high surface tension $\sigma_{\rm lv}\simeq 70\;  {\rm mN m}^{-1}$ (water) and the line tension is as small as~\cite{Schimmele2007} $\tau\simeq 10^{-11}{\rm N}$, we have $R=\tau/\left(\tilde{\tau}\sigma_{\rm lv}\right) \simeq 10^{-10}{\rm m}$.  Therefore, we would need a nanometer-sized pore to observe the wetting and drying.  To observe these morphological transformations in macro- and micro-scale pores, a much larger~\cite{Drelich1996} line tension on the order of $\tau\sim 10^{-5}-10^{-6}{\rm N}$ would be necessary, though such a large value is rejected from recent experimental results~\cite{Bonn2009,Pompe2000,Wang2001,Checco2003,Schimmele2007} and theoretical calculations~\cite{Weijs2011,Schimmele2007,Maheshwari2016}.  However, an ultra-low surface tension, such as the value of  $\sigma_{\rm lv}\simeq 10^{-7} {\rm N/m}$ theoretically predicted for a colloid-polymer mixture~\cite{Vandecan2008}, can also increase the size of the pore  to as much as $R \sim 10^{-5}-10^{-6}{\rm m}$.

The above discussions and our model assume that the line tension $\tau$ is constant and does not depend on the size and the contact angle of the droplet.  However, it has been well recognized for long time that the magnitude of line tension does depend on the size of droplet.  Although a smaller nano-scale droplet has a line tension on the order of~\cite{Schimmele2007}  $\tau\simeq 10^{-11}{\rm N}$, a larger droplet has a larger line tension~\cite{Heim2013}. The line tension can be constant only when the radius of the three-phase contact line is less than $10^{-7}{\rm m}$~\cite{Heim2013}. Furthermore, the line tension should depend not only on the size of droplet but also on the contact angle~\cite{David2007,Heim2013} because intermolecular forces at the three-phase contact line are affected by the geometry at the contact line.  Then, we cannot use the generalized Young's equation, Eq.~(\ref{eq:B15}), which is derived from the variation under the condition of constant line tension~\cite{Schimmele2007} when the line tension depends on the size of the droplet. Therefore, our model cannot be used to describe the size dependence over several orders of magnitude.

\section{\label{sec:sec3}  Volume-dependence of morphology and free energy of a droplet on a concave spherical substrate}

\begin{figure}[htbp]
\begin{center}
\includegraphics[width=0.70\linewidth]{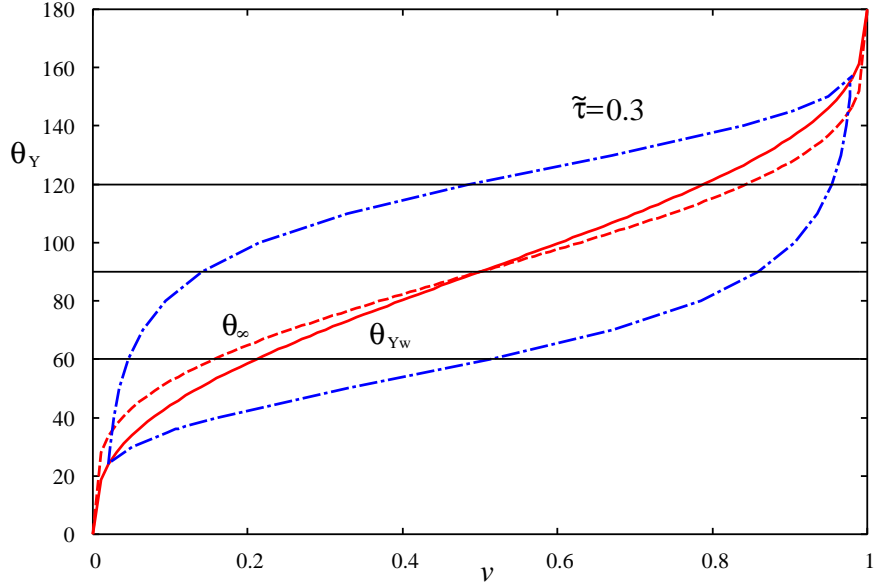}
\caption{
Three routes of changing droplet volume with $\theta_{\rm Y}=60^{\circ}, 90^{\circ}, 120^{\circ}$  (three horizontal lines) when $\tilde{\tau}=0.3$.  When the line crosses the upper curve, the drying transition between a lens-shaped droplet and spherical droplet occurs.  When it crosses the lower curve, the wetting transition between a lens-shaped droplet and a spherical bubble occurs
  }
\label{fig:B5}
\end{center}
\end{figure}

By changing the droplet volume, we can study the volume dependence of the morphology and the free energy of a droplet, which will determine the stability of the superhydrophobic Cassie state~\cite{Whyman2011,Savoy2012,Giacomello2012b,Verho2012,Bormashenko2013b,Papadopoulos2013,Checco2014}.  In Fig.~\ref{fig:B5}, we show three routes with fixed Young's contact angles ($\theta_{\rm Y}=60^{\circ}, 90^{\circ}, 120^{\circ}$) when $\tilde{\tau}=0.3$.  As we increase the dimensionless volume $v$ along one of these lines, the morphology of the droplet changes from a spherical droplet with $\theta_{\rm e}=180^{\circ}$ to a lens-shaped droplet with a convex meniscus, and then to a lens-shaped droplet with a concave meniscus at $\theta_{\rm Y}=\theta_{\infty}$.  Finally, it transforms into a spherical bubble with $\theta_{\rm e}=0^{\circ}$.  The free-energy maximum (barrier) is expected when the line crosses the convex-concave boundary $\theta_{\infty}$ where the size parameter $\rho_{\rm e}$ diverges.

\begin{figure}[htbp]
\begin{center}
\subfigure[]
{
\includegraphics[width=0.5\linewidth]{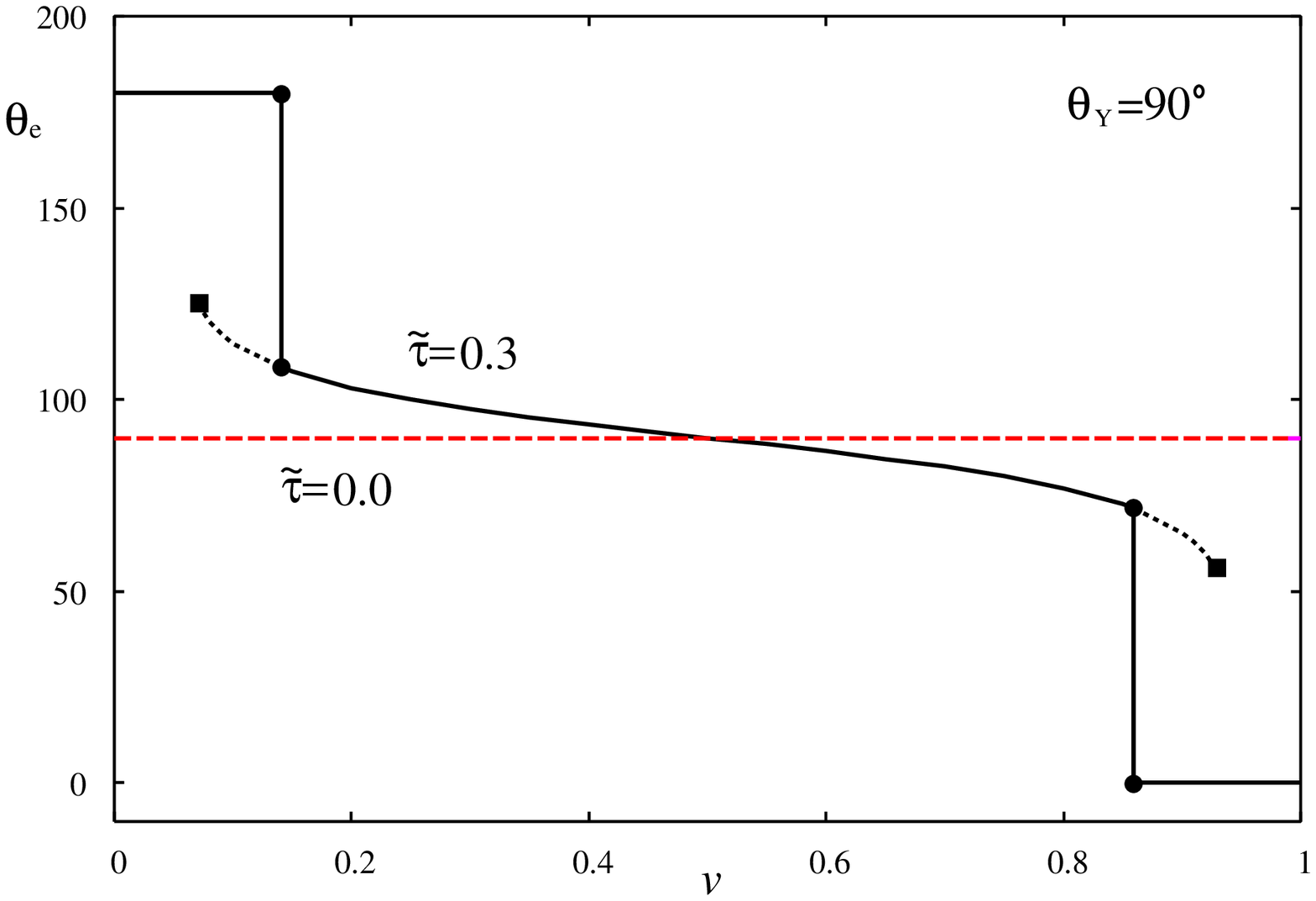}
\label{fig:6a}
}
\subfigure[]
{
\includegraphics[width=0.5\linewidth]{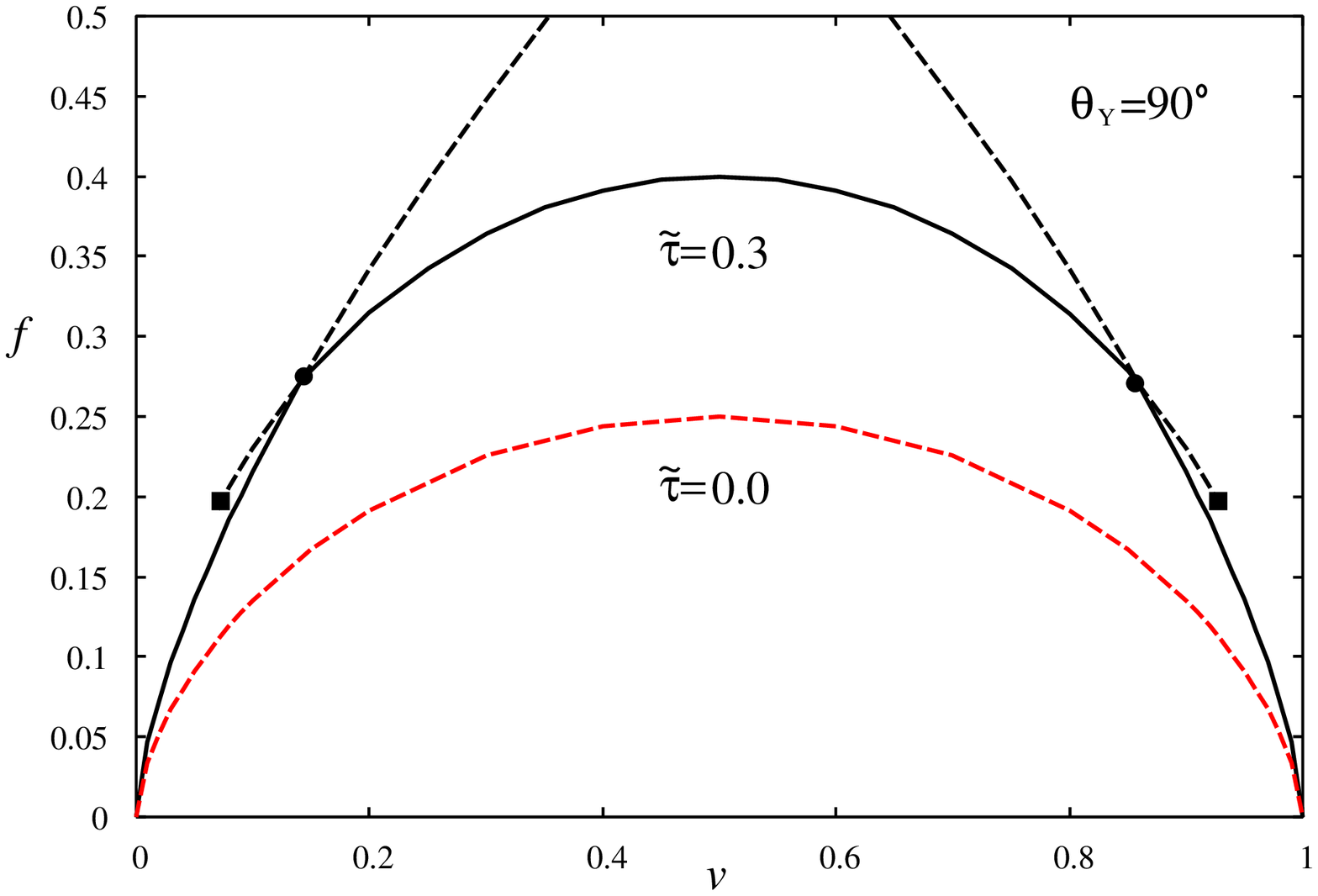}
\label{fig:6b}
}
\end{center}
\caption{
(a) The volume dependence of the equilibrium contact angle $\theta_{\rm e}$ determined from Eq.~(\ref{eq:B16}) when $\theta_{\rm Y}=90^{\circ}$.  The contact angle jumps from $180^{\circ}$ of a spherical droplet to a finite value of a lens-shaped droplet at the drying transition point $v_{\rm dry}\simeq 0.141$, indicated by the filled circles.  It jumps from $0^{\circ}$ of a spherical bubble to a finite value of a lens-shaped droplet at the wetting transition point $v_{\rm wet}\simeq 0.859$, also indicated by the filled circles.  The contact angle for a metastable lens-shaped droplet is shown by the two dashed curves, with filled squares indicating the limit of metastability.  The metastable spherical droplet and bubble are not shown because they have constant angles of $180^{\circ}$ and $90^{\circ}$, respectively.  (b) The volume dependence of the free energy $f$. The drying and wetting points are indicated by filled circles and the stability limits of the metastable droplet are indicated by filled squares. Because the substrate is neutral ($\theta_{\rm Y}=90^{\circ}$), the free energies for a droplet as $v\rightarrow 0$ and that for a bubble as $v\rightarrow 1$ are zero.  The free energies for the metastable droplet and bubble are shown by dashed curves.  The maximum of the free-energy barrier occurs at $v=0.5$ when $\theta=\theta_{\infty}$ (Fig.~\ref{fig:B2}).  We also show the free energy when $\tilde{\tau}=0$ (lower dashed curve).  The curves are all symmetric about $v=0.5$ as the substrate is neutral.
 } 
\label{fig:B6}
\end{figure}

Figure \ref{fig:B6}(a) shows the volume-dependent equilibrium contact angle $\theta_{\rm e}$ when $\tilde{\tau}=0.3$.  The drying transition and the wetting transition occur at $v_{\rm dry}\simeq 0.141$ and $v_{\rm wet}\simeq 0.859$, respectively.  The contact angle of the lens-shaped droplet changes continuously when $0.141\lesssim v \lesssim0.859$.  This lens-shaped droplet can exist as a metastable droplet with varying contact angles, which is indicated by the two dashed curves outside the region of $0.141\lesssim v \lesssim0.859$.  The metastable spherical droplet and spherical bubble in the region of  $0.141\lesssim v \lesssim0.859$ can also exist, but they are not shown because the contact angles are fixed at $180^{\circ}$ and $0^{\circ}$, respectively.  The equilibrium contact angle is larger than the intrinsic Young's contact angle ($\theta_{\rm e}>\theta_{\rm Y}=90^{\circ}$) when the droplet volume is small ($v<0.5$), whereas it is smaller than Young's contact angle ($\theta_{\rm e}<\theta_{\rm Y}$) when the volume is large ($v>0.5$).  The contact angle is equal to the intrinsic Young's contact angle $\theta_{\rm e}=\theta_{\rm Y}=90^{\circ}$ at $v=0.5$, where the contact line coincides with the equator of the spherical cavity.  As explained in our previous publication~\cite{Iwamatsu2016a},  the line tension contribution in the generalized Young's equation vanishes when the contact line coincides with the equator.  Then, the equilibrium contact angle becomes equal to the intrinsic Young's contact angle ($\theta_{\rm e}=\theta_{\rm Y}=90^{\circ}$).

The free-energy barrier for a neutral wall with $\theta_{\rm Y}=90^{^\circ}$ when $\tilde{\tau}=0.3$ is shown in Fig.~\ref{fig:B6}(b).  This Helmholtz free energy barrier is the work which is necessary to fill the empty cavity by liquid.  In other words, this energy is the work necessary to destroy the Cassie superhydrophobic substrate.  The free-energy barrier without line tension ($\tilde{\tau}=0$) is also shown.  The line tension effect on the energy barrier is the most important at the maximum.  The free energy for a completely dried and empty cavity at $v=0$ is given by
\begin{equation}
f_{\rm empty}=f\left(v\rightarrow 0\right)=0
\label{eq:B25}
\end{equation}
from Eq.~(\ref{eq:B4}), whereas that for a completely filled cavity at $v=1$ is given by
\begin{equation}
f_{\rm filled}=f\left(v\rightarrow 1\right)=-\cos\theta_{\rm Y}.
\label{eq:B26}
\end{equation}
Therefore, $f_{\rm filled}<f_{\rm empty}$ when the inner wall of the cavity is hydrophilic, $\theta_{\rm Y}<90^{\circ}$.  Then, the filled configuration will be more stable than the empty configuration.  On the other hand, $f_{\rm filled}>f_{\rm empty}$ when the inner wall of the cavity is hydrophobic, $\theta_{\rm Y}>90^{\circ}$, as expected.  In Fig.~\ref{fig:B6}(b), the inner wall is neutral ($\theta_{\rm Y}=90^{\circ}$), such that $f_{\rm empty}=f_{\rm filled}=0$.  Therefore, a completely dried cavity and a completely filled cavity have the same free energy. 

\begin{figure}[htbp]
\begin{center}
\includegraphics[width=0.7\linewidth]{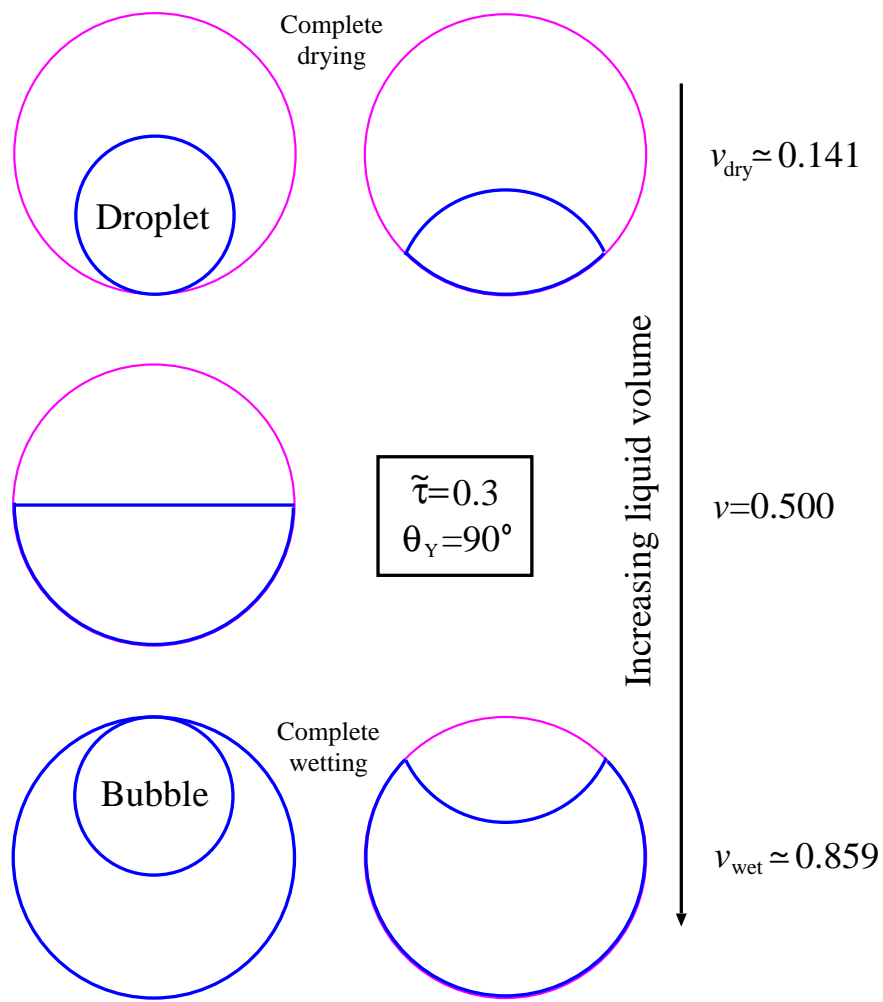}
\caption{
The morphological evolution of a droplet placed on the bottom of a spherical cavity when $\tilde{\tau}=0.3$ as the droplet volume is increased. In this case, the substrate is neutral, with $\theta_{\rm Y}=90^{\circ}$, such that the morphological evolution is symmetric about $\theta_{\rm e}=90^{\circ}$ between the bubble and the droplet.
 }
\label{fig:B7}
\end{center}
\end{figure}

This morphological transition from a spherical droplet to a spherical bubble induced by the volume change has the free-energy barrier to overcome, as shown in Fig.~\ref{fig:B6}(b), whose maximum free energy $f_{\rm Max}$ occurs when the meniscus becomes flat ($\theta_{\rm e}=\theta_{\infty}$ and $\rho_{\rm e}\rightarrow \infty$).  This maximum free energy is given by
\begin{eqnarray}
f_{\rm Max}&=&f\left(\theta=\theta_{\infty}\right) \nonumber \\
&=&\frac{\sin^{2}\theta_{\infty}}{4}
+\frac{\cos\theta_{\rm Y}\left(\cos\theta_{\infty}-1\right)}{2}
+\frac{\tilde{\tau}\sin\theta_{\infty}}{2}
\label{eq:B27}
\end{eqnarray}
from Eq.~(\ref{eq:B4}), where the contact angle $\theta_{\infty}$, which corresponds to the maximum of free energy, satisfies
\begin{equation}
\sin\theta_{\infty}-\cos\theta_{\rm Y}\tan\theta_{\infty}+\tilde{\tau}=0
\label{eq:B28}
\end{equation} 
from Eq.~(\ref{eq:B16}) as $\rho_{\rm e}\rightarrow \infty$.  
These results in Eqs. (27) and (28) agree with those obtained by Bormashenko and Whyman~\cite{Bormashenko2013b}, who used a model in which the droplet meniscus remained flat during the evolution of volume.  Our model, however, assumes that the meniscus changes from convex to concave (Fig.~\ref{fig:B6}(a)) and that the free energy always remains at a minimum.  Our free-energy barrier, shown in Fig.~\ref{fig:B6}(b), is close to the minimum free-energy path~\cite{Sheng2007,Iwamatsu2009} of evolution, where the wetting coordinate~\cite{Sheng2007} is simply given by the dimensionless volume $v$. Therefore, our prediction should be closer to the minimum free-energy path of an actual transition pathway, although we assumed the completely spherical meniscus of a droplet.

As shown in Fig.~\ref{fig:B6}, the line tension cannot be negligible for this free-energy maximum $f_{\rm Max}$.  The same conclusion was previously reached by Bormashenko and Whyman~\cite{Bormashenko2013b}.  In Fig.~\ref{fig:B6}(b), we find from Eq.~(\ref{eq:B27}) that $f_{\rm Max}=0.25+0.5\times0.3=0.4$ for $\tilde{\tau}=0.3$ and $f_{\rm Max}=0.25$ for $\tilde{\tau}=0$ because $\theta_{\rm Y}=\theta_{\infty}=90^{\circ}$.  Therefore, a positive line tension increases the energy barrier $f_{\rm Max}$.  Then, the Cassie-Wenzel transition will be less probable and the superhydrophobic Cassie state will be more stable.  Of course, the effect of negative line tension is the reverse.

The evolution of a droplet shape induced by the volume change is shown in Fig.~\ref{fig:B7}.  The initial droplet shape is spherical sitting on the bottom of the inner wall of a spherical cavity.  As the droplet volume is increased, the spherical droplet transforms into a lens-shaped droplet with a convex meniscus at $v_{\rm dry}\simeq 0.141$ (drying transition).  By further increasing the droplet volume, the meniscus of the droplet changes from convex to concave.  The droplet surface becomes flat when $v=0.5$, where the three-phase contact line coincides with the equator.  Then, the line-tension contribution to the equilibrium contact angle $\theta_{\rm e}$ in Eq.~(\ref{eq:B16}) vanishes.  The equilibrium contact angle is given by the intrinsic Young's contact angle $\theta_{\rm e}=\theta_{\rm Y}=90^{\circ}$. A lens-shaped droplet with a concave meniscus spreads over the whole inner wall of the spherical cavity to form a spherical bubble at $v_{\rm wet}\simeq 0.859$ (wetting transition).   The morphological evolution is symmetric about $\theta_{\rm e}=90^{\circ}$ between the bubble and the droplet because the substrate is neutral with $\theta_{\rm Y}=90^{\circ}$. 

\begin{figure}[htbp]
\begin{center}
\subfigure[]
{
\includegraphics[width=0.5\linewidth]{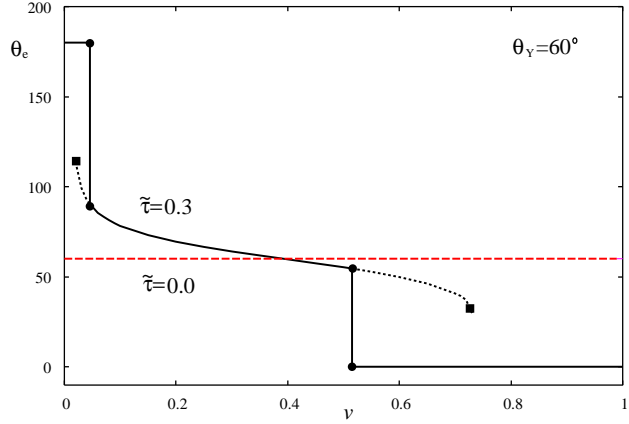}
\label{fig:9a}
}
\subfigure[]
{
\includegraphics[width=0.5\linewidth]{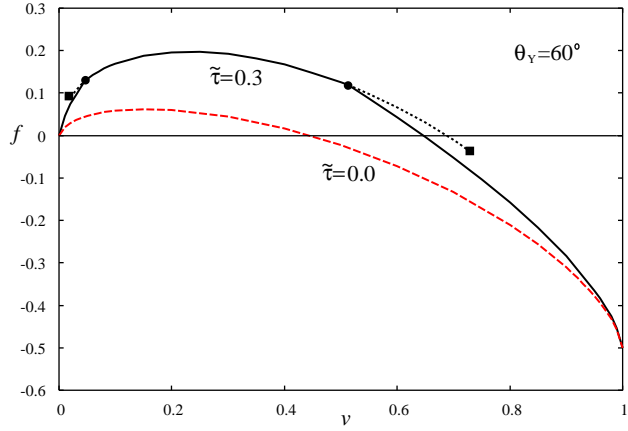}
\label{fig:9b}
}
\end{center}
\caption{
(a) The volume dependence of the equilibrium contact angle determined from Eq.~(\ref{eq:B16}) when the wall is hydrophilic ($\theta_{\rm Y}=60^{\circ}$).  The contact angle jumps from $180^{\circ}$ to a finite value at the drying transition point.  It jumps from $0^{\circ}$ to a finite value at the wetting transition point.  The contact angle for a metastable lens-shaped droplet is shown by a dashed curve.  (b) The volume dependence of the free energy $f$. Because the substrate is hydrophilic ($\theta_{\rm Y}=60^{\circ}$), the free energy of a completely wet cavity, $f\left(v\rightarrow 1\right)$, is lower than that of a completely dry cavity, $f\left(v\rightarrow 0\right)$.  The free energy for the metastable lens-shaped droplet is shown by the dashed curves, but the one for the metastable spherical droplet and spherical bubble, which was shown in Fig.~\ref{fig:B6}(b), is omitted. The maximum of the free-energy barrier occurs at $v_{\infty}\simeq 0.230$ when $\theta=\theta_{\infty}\simeq 67.8^{\circ}$ (Fig.~\ref{fig:B2}).  We also show the free energy when $\tilde{\tau}=0$ (lower dashed curve).
 } 
\label{fig:B8}
\end{figure}

So far, we have considered a neutral wall with $\theta_{\rm Y}=90^{\circ}$.  The result for the hydrophilic wall with $\theta_{\rm Y}=60^{\circ}$ is shown in Fig.~\ref{fig:B8}.  Figure \ref{fig:B8}(a) shows the volume-dependent equilibrium contact angle $\theta_{\rm e}$ when $\tilde{\tau}=0.3$.  Because the inner wall is hydrophilic, the drying and wetting transitions occur at smaller volumes ($v_{\rm dry}\simeq 0.046, v_{\rm wet}\simeq 0.515$).  Therefore, the wall is more easily wetted by the droplet.  

\begin{figure}[htbp]
\begin{center}
\includegraphics[width=0.70\linewidth]{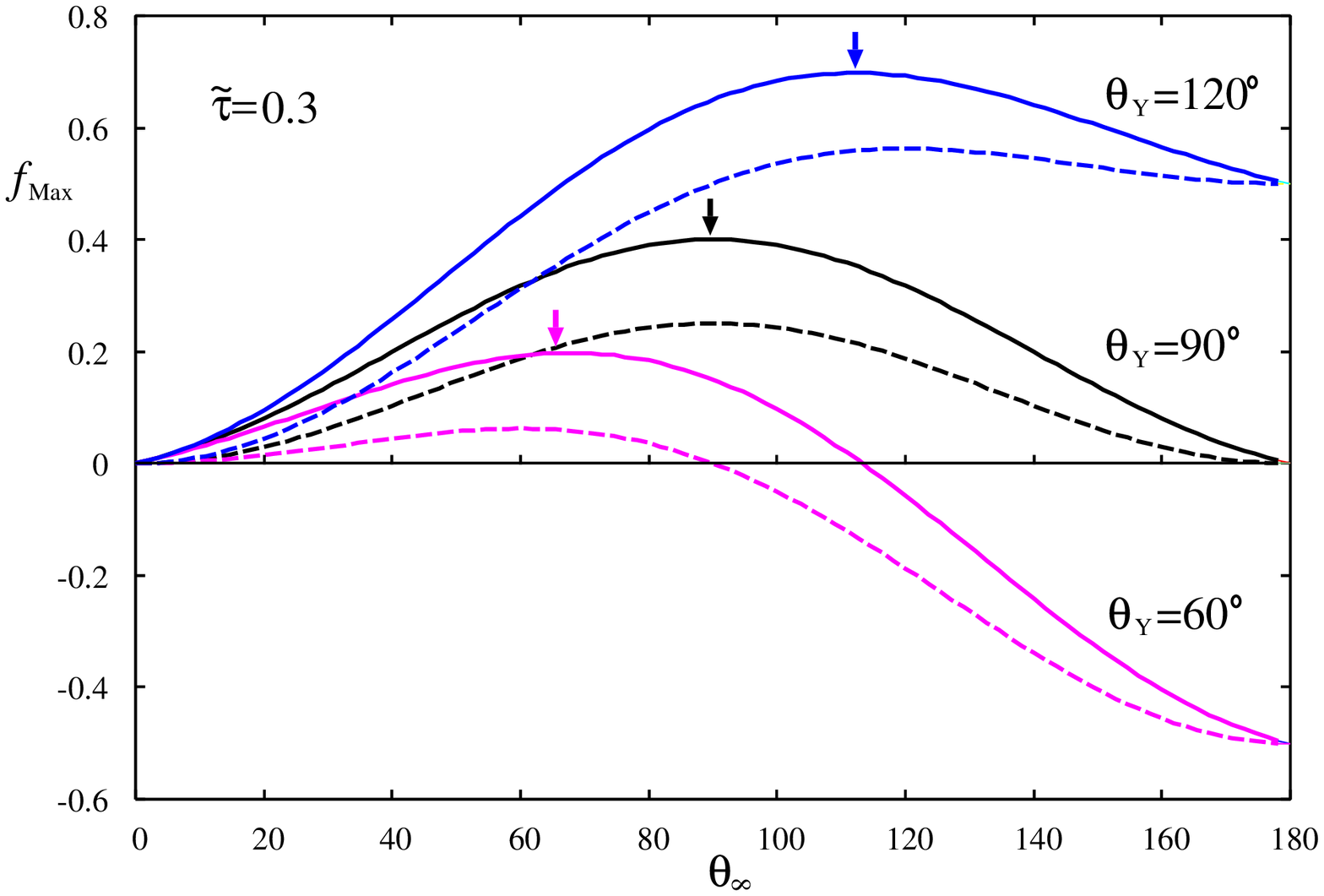}
\caption{
The free-energy maximum $f_{\rm Max}$ given by Eq.~(\ref{eq:B27}) as a function of the contact angle $\theta_{\infty}$, where the meniscus becomes flat.  The contact angle $\theta_{\infty}$ is determined from Eq.~(\ref{eq:B28}), which can be combined with Eq.~(\ref{eq:B1}) to give the volume $v_{\infty}$ where the free-energy maximum occurs.  The three arrows at $\theta_{\infty}\simeq 67.8^{\circ}$, $90^{\circ}$ and $112.2^{\circ}$ indicate the contact angle $\theta_{\infty}$, where the maximum occurs when $\tilde{\tau}=0.3$.
 }
\label{fig:B9}
\end{center}
\end{figure}

The free-energy barrier shown in Fig.~\ref{fig:B8}(b) also indicates that the free energy of a completely filled cavity, $f_{\rm filled}=-\cos\theta_{\rm Y}=-\cos 60^{\circ}=-0.5$, is lower than that of a completely dry cavity,  $f_{\rm empty}=0$.  Therefore, a filled cavity is more favorable than a empty cavity.  The maximum of the free energy $\theta_{\infty}$ shifts to lower volumes in Fig.~\ref{fig:B6}(b).  Furthermore, the free-energy barrier decreases more in a hydrophilic wall than in a neutral wall in Fig.~\ref{fig:B6}(a).

Figure~\ref{fig:B9} shows the maximum of the free-energy barrier $f_{\rm Max}$ in Eq.~(\ref{eq:B27}) as a function of the contact angle $\theta_{\infty}$ at the maximum. 
The contact angle $\theta_{\infty}$ can be determined from Eq.~(\ref{eq:B28}) for the wettability $\theta_{\rm Y}$ and the line tension $\tilde{\tau}$, which can be combined with Eq.~(\ref{eq:B1}) or Fig.~\ref{fig:B1} to determine the volume $v_{\infty}$, where the free-energy maximum is attained.  Figure ~\ref{fig:B9} clearly shows that a positive line tension $\tilde{\tau}=0.3$ increases the barrier significantly, particularly when the wall is hydrophilic ($\theta_{\rm Y}=60^{\circ}$).  Therefore, the line tension can be important to sustain the hydrophobic Cassie state of hydrophilic materials~\cite{Bormashenko2013b}.  On the other hand, its effect is relatively small for the hydrophobic wall because the free-energy barrier $f_{\rm Max}$ without the line tension is already high for the hydrophobic wall.  Therefore, it is energetically unfavorable to fill a hydrophobic cavity ($\theta_{\rm Y}=120^{\circ}$).

\begin{figure}[htbp]
\begin{center}
\subfigure[]
{
\includegraphics[width=0.5\linewidth]{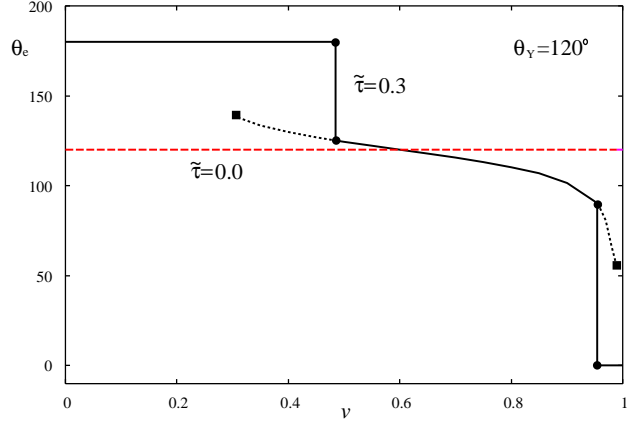}
\label{fig:11a}
}
\subfigure[]
{
\includegraphics[width=0.5\linewidth]{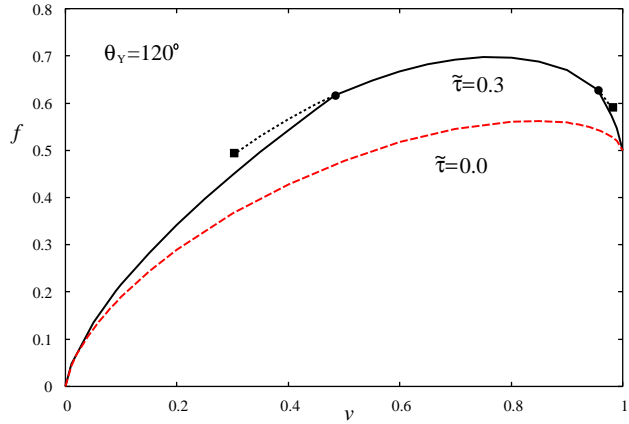}
\label{fig:11b}
}
\end{center}
\caption{
(a) The volume dependence of the equilibrium contact angle determined from Eq.~(\ref{eq:B16}) when the wall is hydrophobic ($\theta_{\rm Y}=120^{\circ}$).  The contact angle for a metastable lens-shaped droplet is shown by a dashed curve.  (b) The volume dependence of the free energy $f$. Because the substrate is hydrophobic, the free energies of a completely wet cavity $f\left(v\rightarrow 1\right)$ are higher than those of a completely dried cavity $f\left(v\rightarrow 0\right)$.  The free energy for the metastable lens-shaped droplet is shown by the dashed curves.  The maximum of the free-energy barrier occurs at $v_{\infty}\simeq 0.770$ when $\theta=\theta_{\infty}\simeq 112.2^{\circ}$ (Fig.~\ref{fig:B2}).  We also show the free energy when $\tilde{\tau}=0$ (lower dashed curve).
 } 
\label{fig:B10}
\end{figure}

In fact, as shown in Fig.~\ref{fig:B10}(b), the free-energy barrier for the hydrophobic wall ($\theta_{\rm Y}=120^{\circ}$)  becomes much higher than the hydrophilic wall (Fig.~\ref{fig:B8}) or the neutral wall (Fig.~\ref{fig:B6}).  Furthermore, the empty cavity is more favorable than the filled cavity ($f_{\rm empty}<f_{\rm filled}$). The effect of line tension becomes relatively weak as compared to the hydrophilic ($\theta_{\rm Y}=60^{\circ}$) and neutral ($\theta_{\rm Y}=90^{\circ}$) walls, but it is still not negligible (Fig.~\ref{fig:B9}).  Naturally, the superhydrophobic Cassie state is more stable for the hydrophobic wall than for the hydrophilic wall~\cite{Whyman2011,Savoy2012,Giacomello2012b,Verho2012,Bormashenko2013b,Papadopoulos2013,Checco2014}.

The volume-dependent contact angle for the hydrophobic wall in Fig.~\ref{fig:B10}(a) shows that the drying transition and the wetting transition occur at larger volumes $v_{\rm dry}\simeq 0.485, v_{\rm wet}\simeq 0.954$.  The cavity wall stays dry until the droplet volume becomes relatively large because the wall is hydrophobic.

As shown in our previous paper~\cite{Iwamatsu2016a}, the effect of negative line tension is totally different from that of positive line tension.  Neither the drying transition nor the wetting transition appear.  The equilibrium contact angle $\theta_{\rm e}$ depends weakly on the volume of the droplet and remains near the Young's contact angle, $\theta_{\rm Y}$.  However, the effect of line tension on the free-energy barrier can be significant from Eq.~(\ref{eq:B27}).  The maximum energy $f_{\rm Max}$ will be reduced for negative line tension.  Then, the energy barrier of filling a spherical pore will be reduced.

In the previous section, we discussed that the line tension must be positive and large in order to observe the line-tension induced drying and wetting within a cavity. Furthermore, not only the line tension but also the contact angle hysteresis, which is always present on real surfaces~\cite{Bormashenko2013a} will also affect these transitions.  Then, the positions of the drying and wetting transitions and the barrier maximum when the liquid is injected would be different than those when the liquid is extracted.   

Only a small number of experimental study exists for the wetting of surfaces made of spherical cavities~\cite{Abdelsalam2005,Lloyd2015}.  However, those studies payed most attention to the contact-angle hysteresis of a macroscopic droplet, and did not investigate the wetting and Cassie-Wenzel transition of individual cavity.  The Cassie-Wenzel transition has been studied experimentally using cavities with simpler geometry~\cite{Papadopoulos2013,Checco2014}.  For example, Papadopoulos et al.~\cite{Papadopoulos2013} studied the breakdown of the hydrophobic Cassie state by the evaporation of a macroscopic droplet.  The depinning and retraction of the three-phase contact line by evaporation and the subsequent invasion of cavities by liquid (Cassie-Wenzel transition) were directly observed by confocal microscopy. The Cassie-Wenzel transition takes place by the meniscus invading the cavity progressively by increasing the droplet volume inside the cavity~\cite{Whyman2011,Savoy2012,Giacomello2012b,Papadopoulos2013,Checco2014}. Therefore, the free-energy barrier must be crossed during the course of invasion. However, no information about the free-energy barrier of the transition was reported.  

The Cassie-Wenzel transition has also been studied by applying the pressure to the liquid of droplet~\cite{Checco2014, Giacomello2012b}.  In this case, we have to study the Gibbs free energy~\cite{Iwamatsu2015a} $G=F-\Delta p V$ instead of the Helmholtz free energy $F$ in Eq.~(\ref{eq:B3}), where $\Delta p$ is the applied pressure.  The maximum of the free energy does not occurs when the meniscus becomes flat ($r\rightarrow\infty$).  Instead, it occurs when the radius of the liquid-vapor interface $r$ is given by the Laplace formula~\cite{Iwamatsu2015a} $r=2\sigma_{\rm lv}/\Delta p$, and the free-energy barrier is lowered by the amount $-\Delta p V$ from that given by Eq.~(\ref{eq:B27}).  Then, the Cassie-Wenzel transition will be accelerated by applying the pressure $\Delta p>0$.  Numerically, our result of the free-energy barrier $f_{\rm Max}\simeq 0.1-0.5$ is the same order of magnitude as that obtained for simpler geometry~\cite{Checco2014,Giacomello2012b}.  In fact, the free-energy barrier in our spherical-cavity model will be $F_{\rm Max}=4\pi R^{2}\sigma_{\rm lv}f_{\rm Max}\simeq \left(1.4\times10^{-9}{\rm J}\right)\times f_{\rm Max}$ for the water droplet ($\sigma_{\rm lv}=70\;{\rm mNm}^{-1}$) in a micro-scale pore~\cite{Papadopoulos2013} with $R=40 \mu{\rm m}$ and $F_{\rm Max}\simeq \left(1.4\times10^{-15}{\rm J}\right)\times f_{\rm Max}$ in a nano-scale pore~\cite{Checco2014} with $R=40 {\rm nm}$, which would be difficult to detect.

\section{\label{sec:sec5}Conclusion}

In this study, we considered the volume dependence of the morphology and free energy of a lens-shaped droplet placed on the bottom of a spherical cavity within the capillary model.  The line-tension effects were included, and they were expressed by a scaled line tension. Then, the contact angle was determined from the generalized Young's equation, which took into account the effects of line tension.  The morphology and the free energy were studied using the mathematically rigorous formula for the free energy~\cite{Iwamatsu2016a}. The droplet morphology changed from a spherical droplet to a lens-shaped droplet with a convex meniscus, then, to that with a concave meniscus as the droplet volume increased.  Finally, the lens-shaped droplet spread over the whole inner wall of the cavity to leave a spherical bubble.  The free energy showed a barrier as a function of the droplet volume whose maximum occurred when the meniscus became flat.  This free-energy maximum was given by an analytic formula, which showed that the line-tension contribution was largest at the maximum and could not be treated as a small parameter.

In conclusion, we studied the free-energy barrier of filling a spherical cavity, including the line-tension effect.  The line-tension contribution can be important in determining the maximum of the energy barrier.  The positive line tension increases the energy barrier. Thus, the Cassie-to-Wenzel transition is less probable and the superhydrophobic Cassie state is more stable.   In this study, we used the simplest capillary model and neglected the noncircular fluctuation of the contact line~\cite{Dobbs1999}.  The capillary model will be less reliable if the liquid-substrate interaction is long-ranged represented by the disjoining pressure~\cite{Schimmele2007}.  The noncircular fluctuation will cause instability of the lens-shaped droplet if the line tension is negative, although the higher-order contribution~\cite{Berg2010} will stabilize the fluctuation.

\begin{acknowledgement}
The author is grateful to Professor Edward Bormashenko (Ariel University, Israel) for suggesting the problem of the Cassie-Wenzel stability and for his useful comments on the manuscript.
\end{acknowledgement}




\begin{tocentry}

\begin{center}
\includegraphics[width=0.55\linewidth]{figb8b.eps}
\end{center}
\end{tocentry}

\end{document}